\documentclass[runningheads]{llncs}
\usepackage{graphicx} 
\usepackage[T1]{fontenc}
\usepackage{mwe} 
\usepackage{multirow}
\usepackage{booktabs}
\usepackage{amsmath}
\usepackage{hyperref}
\usepackage{amssymb}

\begin{document}

\title{Registration of Longitudinal Liver Examinations \\ for Tumor Progress Assessment}
\titlerunning{Registration of Longitudinal Liver Examinations}

\author{Walid Yassine\inst{1, 2}\index{Yassine, Walid}\and
Martin Charachon\inst{2}\index{Charachon, Martin} \and
Céline Hudelot\inst{1}\index{Hudelot, Céline} \and
Roberto Ardon\inst{2}\index{Ardon, Roberto}
}
\authorrunning{W. Yassine et al.}

\institute{MICS, CentraleSupélec, Université Paris Saclay - France\\
\email{fname.lname@centralesupelec.fr} \\ \and
Incepto Medical - France \\ 
\email{fname.lname@incepto-medical.com}
}

\maketitle        

\begin{abstract}
Assessing cancer progression in liver CT scans is a clinical challenge, requiring a comparison of scans at different times for the same patient. 
Practitioners must identify existing tumors, compare them with prior exams, identify new tumors, and evaluate overall disease evolution. 
This process is particularly complex in liver examinations due to misalignment between exams caused by several factors. Indeed, longitudinal liver examinations can undergo different non-pathological and pathological changes due to non-rigid deformations, the appearance or disappearance of pathologies, and other variations. 
In such cases, existing registration approaches, mainly based on intrinsic features may distort tumor regions, biasing the tumor progress evaluation step and the corresponding diagnosis. 
This work proposes a registration method based only on geometrical and anatomical information from liver segmentation, aimed at aligning longitudinal liver images for aided diagnosis.
The proposed method is trained and tested on longitudinal liver CT scans, with 317 patients for training and 53 for testing.
Our experimental results support our claims by showing that our method is better than other registration techniques by providing a smoother deformation while preserving the tumor burden\footnote{Total volume of tissues considered as tumor.} within the volume. Qualitative results emphasize the importance of smooth deformations in preserving tumor appearance.


\end{abstract}

\section{Introduction}\label{Introduction}

\begin{figure}[h!]
\includegraphics[width=0.9\textwidth]
{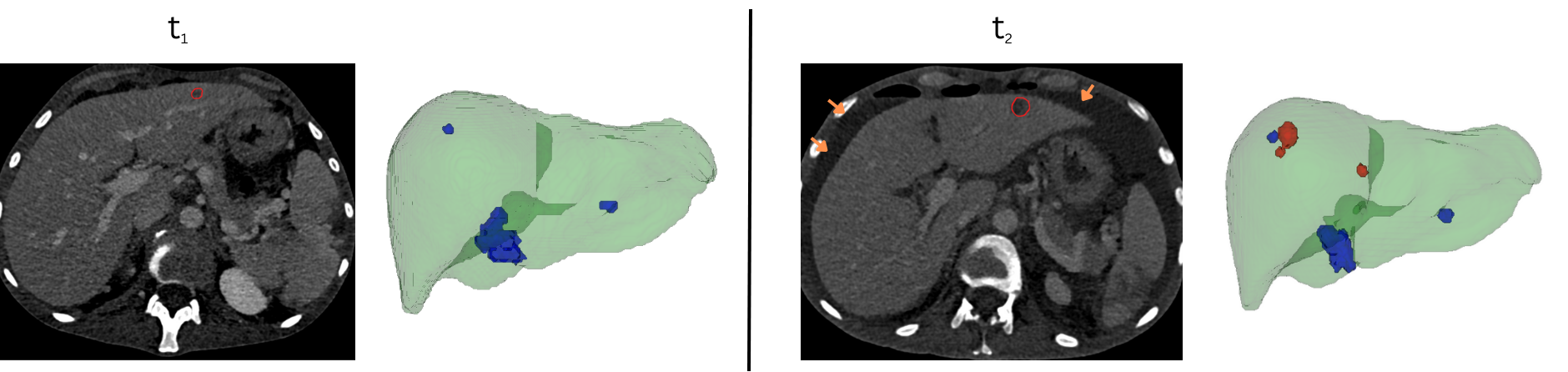}
\centering
\caption{Liver longitudinal exams: The images show a growing lesion in red and reveal changes in the liver appearance after two months, e.g., effusion around the liver (orange arrows). The liver segmentation mask is presented in green, existing tumors in blue, and new tumors in red.}
\label{fig:intro}
\end{figure}

Evaluating cancer progression in liver CT scans during patient follow-up poses a significant clinical challenge. 
In the healthcare workflow, practitioners compare scans taken at different times for the same patient. 
This process involves identifying new and pre-existing lesions in the latest scans and assessing tumor progression according to the RECIST \cite{RECIST} guidelines. 
This monitoring demands considerable effort as practitioners must recognize previously detected lesions, compare them to their counterparts in prior exams, identify any newly appearing tumors, and evaluate the overall evolution of the disease. 
This process gets more complicated in liver examinations due to misalignment and variations caused by temporal factors. 
In longitudinal studies, variations such as patient movements, positioning, and organ displacement are common. 
Moreover, being a non-rigid organ, the liver presents additional complexities with recurring changes like large deformations due to stomach pressure, alterations in pathology size (cancerous or not), increased fat content, changes in vessel size, and bile duct dilation. 
Fig.~\ref{fig:intro} illustrates the complexity of the process by representing a longitudinal exam with multiple lesions. 
Tools performing automatic alignment of CT scans for precise tumor follow-up may reduce radiologists mental burden.

\paragraph{\textbf{Image Registration for longitudinal studies}}

Registration \footnote{The terms 'alignment' and 'registration' are used interchangeably throughout the document.} of medical images has garnered significant attention in the scientific literature. 
As in many other domains, numerous approaches based on deep neural networks have been developed 
\cite{Voxelmorph,Voxelmorph++,ConditionalDeformable-Unsupervised,Diffeomorphe,CycleMorph,Unsupervised-Abdominal,Unsupervised-Transformer}. These approaches involve feeding pairs of images (the moving and the fixed image) into a neural network. 
The network predicts a displacement field to align the moving image with the fixed image. 
In the context of tumor progression monitoring, such registration methods may significantly impact the monitoring process. For example, non-rigid registration of a liver may distort tumor regions when comparing a liver with a tumor at time $t$ to the same liver at $t + 1$ with a larger tumor.
Intrinsic organ characteristics, like rigidity, also impact the registration process. 
Existing literature primarily addresses the registration problem by exploiting the image content. 
In the case of liver exams, many pathological and non-pathological variations can occur. 
Registration based on intrinsic features (e.g., perceptual content) can impact internal liver structures and, 
consequently, the tumor burden, affecting the tumor progression evaluation.
To our knowledge, only a few studies have explored tumor change/evolution detection in longitudinal images. Existing works mainly focus on multiple sclerosis progression in brain MRI \cite{BrainMS-1,BrainMS-3,BrainMS-0,BrainMS-2}. 
Nevertheless, the brain is not subject to the same factors as the liver. This raises concerns about the applicability of these approaches to content-based liver registration if one aims at preserving post-registration tumor integrity.
Some methods use the predicted displacement field to identify regions of tumor changes by detecting warping 
within the brain \cite{Brain-deformation}. Such approaches are considered unsuitable in liver examinations due to 
the organ's non-rigidity, the diverse pathological/non-pathological changes within the liver, and temporal variations between exams.
Alternatively, some approaches adopt a two-step process for this monitoring task: 
Independent tumor detection, followed by image registration to deduce tumor correspondence in time \cite{Lesion_Tracking_1,Lesion_Tracking_2,Lesion_Tracking_3}. 
\cite{Brain-Longitudinal-Motivation} highlighted that independent tumor detection can be less precise and sensitive than considering both exams simultaneously, aligning more closely with real clinical contexts.

\paragraph{\textbf{Medical Image Registration Background}} A recent review \cite{Survey-Registration} evaluates various unsupervised 3D medical image registration methods. 
The VoxelMorph architecture \cite{Voxelmorph} has been widely employed. It predicts a displacement field coupled with a differentiable spatial transformation layer (STN) \cite{STN} to apply transformations. 
\cite{Diffeomorphe} proposes a bidirectional diffeomorphic registration, introducing inverse coherence and anti-folding losses for displacement fields in both forward and backward directions. 
More recent approaches have introduced transformer-based networks \cite{Unsupervised-Transformer}, capitalizing on the ability of transformer blocks to capture global image features during registration rather than relying solely on local information. 
\cite{CycleMorph} introduced Cyclemorph, a cyclic registration method incorporating cyclic consistency into the network's loss function to enhance performance. 
Incremental transformations are also discussed in the literature. 
\cite{Neural-ODE,Neural-ODE-2} present a network based on neural ordinary differential equations, aiming to perform iterative registration through a neural network. 
This approach decomposes the displacement field into multiple small steps, each constrained to ensure displacement field regularity. 
Most of these methods rely on image content (or image + organ segmentation) for registration. \\

\textbf{Our main contribution} is to propose a registration framework that aligns longitudinal images based only on geometric and anatomical information from liver segmentation, smoothly extrapolating the displacement field within the liver. 
The rationale behind this is that even though internal structures may not be perfectly aligned, their shape will undergo small deformations while being brought much closer to their counterpart in the reference image. Following the works of \cite{BrainMS-1,BrainMS-3,BrainMS-0,BrainMS-2}, we adhere to the strategy where both images should be considered to design a tumor progression/change detector; this paper proposes an adapted registration method to such frameworks.

\section{Proposed Methodology}\label{Methodology}

\begin{figure}[h]
\includegraphics[width=\textwidth]
{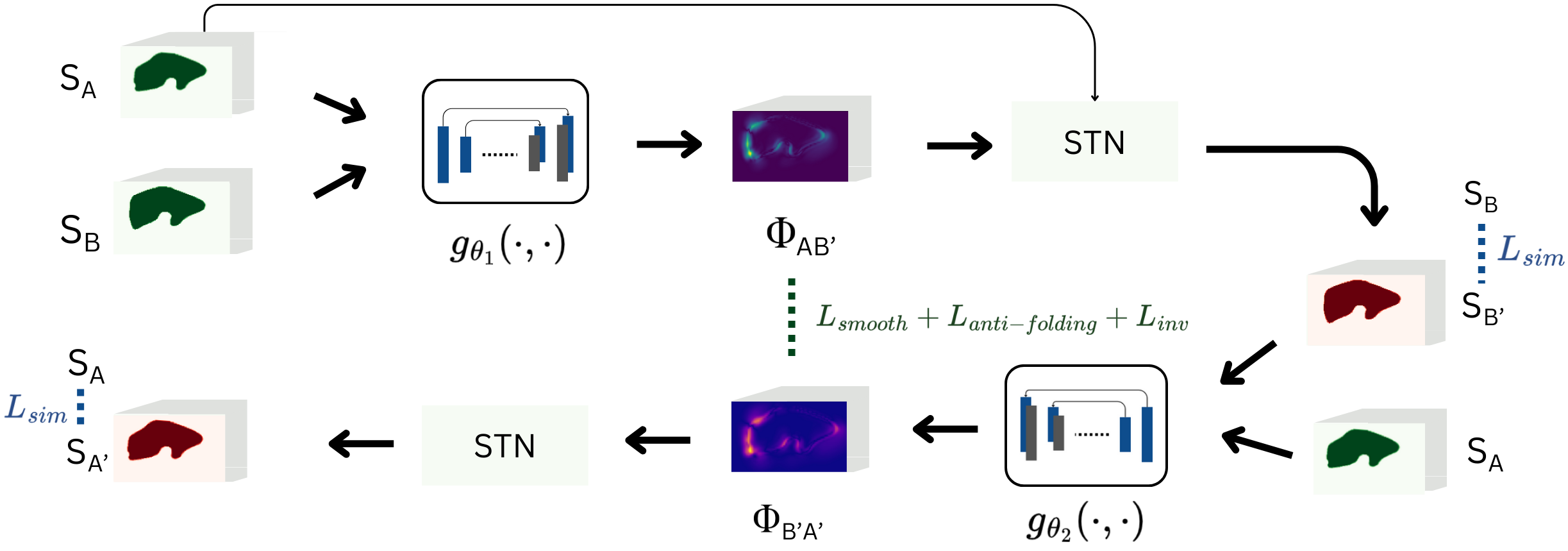}
\centering
\caption{\textbf{Cyclic Diffeomorphic Registration:} $g_{\theta_1}$ takes as input the moving and fixed segmentation masks $S_A$ and $S_B$ and generates a displacement field $\phi_{AB'}$. This field is applied to $S_A$ to obtain $S_{B'}$, the aligned segmentation mask. $S_{B'}$ goes through $g_{\theta_2}$ (with $S_A$) to obtain the cyclically transformed segmentation $S_{A'}$.}
\label{fig:methodology}
\end{figure}

Let $A$ and $B$ represent images of the same patient captured at different time points, denoting the moving and fixed images, respectively. $A$ and $B$ are affinely aligned during a pre-processing step. 
We assume that we are equipped with a liver segmentation tool, and we note $S_A$ and $S_B$ as the 3D liver segmentation masks in images $A$ and $B$, respectively. 
These entities are defined in a three-dimensional image space $\Omega \in \mathbb{R}^3$. 
Our objective is to determine a transformation function that aligns the segmentation $S_A$ of the moving image with the segmentation $S_B$ of the fixed image.
Inspired by the previously mentioned works, we propose a new registration framework focusing on segmentation maps, illustrated in Fig.~\ref{fig:methodology}.
The framework trains a model $g_{\theta_1}$ to generate a displacement field $\phi_{AB'}$ when provided with a pair of segmentations $S_A$ and $S_B$. 
Each component of $\phi_{AB'}$ is a three-dimensional vector indicating the displacement of a specific voxel. Subsequently, we use a differentiable operation based on a spatial transformer network (STN) to apply this displacement field $\phi_{AB'}$ to the segmentation mask $S_A$, resulting in $S_B'$: the segmentation mask aligned with $S_B$. To ensure effective registration, it is necessary to impose a set of constraints:

\paragraph{\textbf{Displacement Field Regularity}}
This first classical constraint penalizes the norm of the derivatives of the displacement field to ensure its local smoothness. 

\begin{align}\label{eq:eq-smooth}
   L_{\text{smooth}}(\phi) = ||\nabla \phi||_2^2 
   =  \Sigma_{i, j}{ \frac{\partial \phi^i}{\partial x_j}^2}
\end{align}

\paragraph{\textbf{Segmentation alignment}}

A second constraint, $L_{\text{sim}}$, ensures alignment between the transformed mask $S_{B'} = S_A + \phi_{AB'}(S_A)$ and the fixed mask $S_B$.
\begin{align}\label{eq:eq-lsim}
   L_{\text{sim}}(S_{B}, S_{B'})  = 1 - \text{\textit{DSC}}(S_{B}, S_{B'}) 
\end{align}

where, \textit{DSC} refers here and henceforth to the Dice similarity coefficient \cite{Dice}.

\noindent Two additional constraints, inspired by \cite{Diffeomorphe,MotivationDiffeo-1,MotivationDiffeo-2}, are also incorporated to prevent a trivial solution that excessively stretches the contours of $S_A$ in a non-plausible way to fit $S_B$.

\paragraph{\textbf{Anti-Folding}}

The constraint $L_{anti-folding}$, detailed in Eq.~\ref{eq:eq-anti-folding}, prevents displacement field folding, avoiding unrealistic distortions in the image. 
It maintains local smoothness by ensuring non-overlapping gradient directions around each point $p$ in the image space $\Omega$.
\begin{align}\label{eq:eq-anti-folding}
    L_{\text{anti-folding}}(\phi) =  \sum_{\substack{p \in \Omega}} \sum_{i} &\delta \left( \frac{\partial \phi^i}{\partial x_i}(p) + 1 \right) \left|  \frac{\partial \phi^i}{\partial x_i}(p) \right|^2
\end{align}
where, $\delta(Q)$ is an indexing function that penalizes the gradient $\phi$ at folding points. 
Specifically, if $Q \leq 0$, then $\delta(Q) = 1$, otherwise, $\delta(Q) = 0$.

\paragraph{\textbf{Inverse Consistency}}\label{paragraph:inverse-coherence}

The inverse consistency constraint $L_{inv}$, detailed in Eq.~\ref{eq:eq-inverse-loss}, ensures that the predicted displacement fields are invertible and inversely consistent. Following \cite{Diffeomorphe}'s bidirectional approach, forward displacement fields from A to B ($\phi_{AB}$) and from B to A ($\phi_{BA}$), along with their estimated inverse fields, are computed. 
The constraint minimizes the Frobenius norm between the estimated inverse field $\tilde{\phi}_{BA}$ of the forward displacement field $\phi_{AB}$ and the true backward displacement field $\phi_{BA}$ as expressed in Eq.~\ref{eq:eq-inverse-loss}. 
$\tilde{\phi}_{BA}$ is obtained through a sampling function $\zeta$ that operates on the displacement fields. 
More details on the inverse consistency and the anti-folding losses are presented by \cite{Diffeomorphe}.

\begin{align}\label{eq:eq-inverse-loss}
    L_{\text{inv}}(\phi_{\text{AB}}) &= \| \phi_{\text{AB}} - \tilde{\phi}_{\text{AB}} \|_{F}^{2} \quad \text{with} \quad \tilde{\phi}_{\text{AB}} = \zeta(-\phi_{\text{BA}}, \phi_{\text{AB}})
\end{align}

However, the inverse consistency constraint has limitations as there is no guarantee that $\phi_{AB}(S_A) = S_B$, except in cases where $L_{sim} = 0$. 
Consequently, the inverse consistency term can never be zero unless the transformed segmentation mask is identical to $S_B$. 
To illustrate this, consider two points, $a$ and $b$, on the segmentation masks $S_A$ and $S_B$, respectively. 
In response, inspired by \cite{CycleMorph}, we favor a cyclic method over a bidirectional one to compute the displacement fields. For a cyclic path, we obtain two successive displacement fields: one transforms $a$ to $b'$ and the other $b'$ to $a'$. 
The second field $\phi_{b'\rightarrow a'}$ starts where the first $\phi_{a \rightarrow b'}$ ends, ensuring a well-defined inverse consistency loss. By doing this, we enforce the displacement field $\phi_{a \rightarrow b'}$ to be inversely consistent while gradually aligning $b'$ to $b$ during model training. 
Appendix. \ref{appendix:cyclic} provides a more detailed analysis of the motivation behind favoring a cyclic approach over a bidirectional one.
Finally, the total loss function is formulated as follows:

\begin{align}\label{eq:eq-loss-complete}
    L(S_{A}, S_{B}) \hspace{2pt} =  &\alpha \hspace{2pt} L_{\text{sim}}(S_{B'}, S_{B})  + \beta \hspace{2pt} (L_{\text{smooth}}(\phi_{AB'}) + L_{\text{smooth}}(\phi_{B'A'}))  \hspace{2pt} + \notag \\
    & \gamma \hspace{2pt} (L_{\text{anti-folding}}(\phi_{AB'}) + L_{\text{anti-folding}}(\phi_{B'A'})) \notag \\
    & + \mu \hspace{2pt} (L_{\text{inv}}(\phi_{AB'}) + L_{\text{inv}}(\phi_{B'A'}))
\end{align}

\paragraph{\textbf{Incremental Cyclic Diffeomorphic Registration}}

\begin{figure}[h!]
\includegraphics[width=\textwidth]
{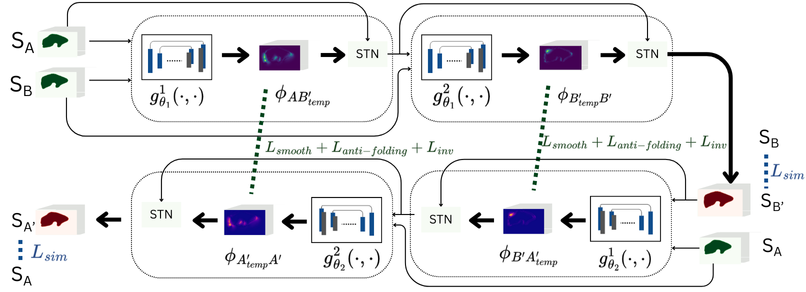}
\centering
\caption{\textbf{Incremental Cyclic Diffeomorphic Registration Framework:} The model takes as input $S_A$ and $S_B$ and generates two displacement fields, $\phi_{AB'_{temp}}$ and $\phi_{B'_{temp}B'}$, to produce the aligned segmentation $S_{B'}$. $S_{B'}$ then goes through the backward path (with $S_A$) to obtain the cyclically transformed segmentation $S_{A'}$.}
\label{fig:methodology-n2}
\end{figure}

We propose to extend our approach by introducing a two-step incremental cyclic diffeomorphic registration process (Fig.~\ref{fig:methodology-n2}). 
The model predicts two displacement fields for each direction. 
The inverse consistency loss imposes constraints to ensure that the displacement fields are inversely consistent.
This approach is motivated by the nature of deformations observed in liver examinations, particularly under external pressure from the stomach, which can induce significant deformations. 
Decomposing the displacement field into two distinct fields reduces the risk of applying large deformations and minimizes potential irregularities in the transformed image. 
During inference, the predicted displacement fields (forward direction) are used to compute aligned versions of $A$ and $S_A$ for the moving image and its segmentation.

\section{Experiments}
\paragraph{\textbf{Dataset and Implementation Details}} 

Our dataset includes 772 pairs of longitudinal liver examinations from 337 patients. 90\% of these patients are allocated for training the registration methods, while a test set of 33 patients is reserved for assessing registration performance. An additional clinical dataset containing 40 exams from 20 liver cancer patients is used to evaluate tumor-wise metrics.
A radiologist manually annotated the tumor masks of this dataset.
The datasets were collected under the GDPR\footnote{General Data Protection Regulation} through a collaboration with a hospital.
An initial liver area cropping and an affine registration (similar to \cite{STN-use}) are applied to eliminate global transformations, particularly in cases involving significant changes in patient position between examinations. 
This aligns with the framework described in Section \ref{Methodology}, where we assumed that the images are affinely aligned, minimizing deformations of large amplitudes. 
For affine registration, a fully convolutional network (FCN) is used to predict the transformation matrix applied to the moving images. 
All Images are resampled to a size of (160, 160, 100), with an average resolution of (1.5, 1.37, 2) mm on the (x, y, z) axes. 
Segmentation masks are generated by a UNet \cite{Unet} model trained on 1000 annotated liver masks, with a DSC of 0.96 (internal test dataset) and 0.971 (LiTS dataset \cite{LiTS}). 
Each CNN network $g^i_{\theta_j}$ for i=\{1,2\} and j=\{1,2\} has a UNet architecture, with four down/up-sampling blocks (the number of filters starts at 8), each containing two convolutional blocks.  
Convolutional blocks include normalization, leaky ReLU activation, and a convolutional layer.
For the incremental framework, we perform two transformation steps, resulting in two fields in each of the forward and backward paths. 
The $g^i_{\theta_j}$ networks do not share their weights (experiments were less conclusive in a weight-sharing setting).
Each network generates a vector field at half the input image resolution, which is then integrated and upsampled to obtain a displacement field at the original resolution.
Three-fold cross-validation experiments (on NVIDIA T4 GPUs) were conducted with an Adam optimizer until convergence and with $\alpha$=1, $\beta$=0.8, $\gamma$=1 and $\mu$=0.4.

\paragraph{\textbf{Evaluation Metrics:}}

i) Alignment between the transformed mask $S_{B'}$ and the fixed mask $S_{B}$ is assessed using the DSC. 
ii) Image content-based coherence is evaluated using normalized cross-correlation (NCC) and mutual information (MI).
iii) Regularity of the displacement field is evaluated using the Jacobian matrix $J_\phi(p) = \nabla\phi(p) = [\partial \phi^i / \partial x_j]_{i,j}$ through its $L_2$ norm and by counting the number of voxels within the liver where the determinant $|J_\phi(p)| \leq 0$ (indicating a non-diffeomorphic field). 
iv) Diffeomorphism is quantified through the $L_1$ distance between the moving image $A$ and its cyclically reconstructed version $A'$. \\

We evaluate our method performance against different registration frameworks: Nifty Reg \cite{Nifty_Reg} (cubic B-splines); 
VoxelMorph under two configurations: 1) trained on images and segmentations with a loss based on image similarity (NCC) and segmentation similarity (DSC), and 2) on segmentations only.
Under both configurations, a regularity loss on the displacement field is employed as in Eq.~\ref{eq:eq-smooth}. We also evaluate the incremental framework coupled with the diffeomorphic parametrization (Diffeo\_inc2), the cyclic diffeomorphic framework with and without incremetal steps (DiffeoCyc\_inc-2, DiffeoCyc\_inc-1 resp.).

To our knowledge, no public dataset exists for longitudinal liver examinations (more details in Appendix \ref{appendix:literature}). 
The difficulty in acquiring such datasets stems from the temporal link between exams.
This limitation prevents benchmarking on public datasets in this specific context.

For tumor-related metrics, evaluation is done on the clinical dataset of 20 patients with 30 annotated tumor masks.
We consider only tumors persisting between examinations, whether stable, growing, or shrinking.
We evaluate the number of matched lesions, the mean tumor inclusion ratio post-registration (overlap of segmentations over the real tumor volume), and the relative error in tumor burden by comparing the tumor burden within an exam pre/post-registration.
A tumor is matched if its inclusion ratio is above 10\%. 
Results significance has been tested with permutation-based statistical tests \cite{Permutation_tests}.

\section{Results and Discussion}
\paragraph{\textbf{Results}}
For field regularity metrics, DiffeoCyc\_inc-2 exhibits values of 0.002, 0.008, and 1 for $||\nabla\phi||_2^2$,  $||A - A'||_1$, and $|J \leq 0|$ metrics respectively. In contrast, NiftyReg (resp. Voxelmorph (NCC + DSC)), which are content-based methods, report values of 0.09 (resp. 0.04), 0.01 (resp. 0.027), 45145 (resp. 12845). The difference between DiffeoCyc\_inc-2 and NiftyReg (resp. Voxelmorph) is significant ($p_{\text{value}}<0.01$).
NCC and MI metrics report comparable values ($p_{\text{value}} > 0.01$) across methods between 0.43 and 0.45, with 0.37 for Voxelmorph (NCC + DSC). 
For tumor-related metrics, NiftyReg manages to match 1 more tumor than the other methods (tumor inclusion ratio = 0.58). The tumor burden relative error for all methods varies between 0.11 and 0.16. DiffeoCyc\_inc-2 reports a displacement field regularity of 0.002 and 0 voxels with non-diffeomorphic deformations.

\paragraph{\textbf{Discussion}}

\begin{figure}[h!]
\includegraphics[width=0.93\textwidth]
{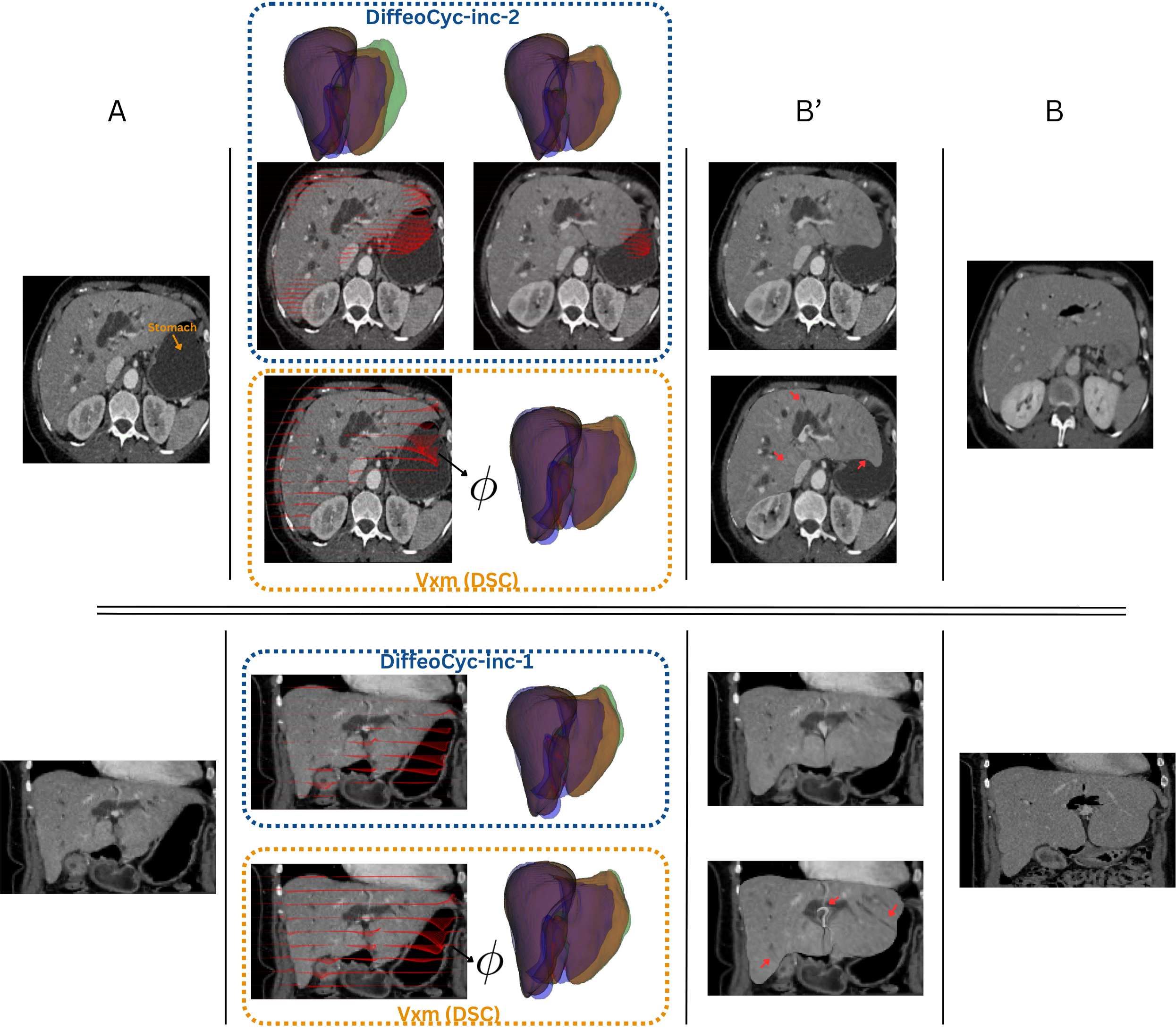}
\centering
\caption{Left to right: Moving image $A$, displacement field $\phi$ (in red), transformed image $B'$, and fixed image $B$. 3D liver masks are presented in blue for A, red for B', and green for B. Red arrows highlight unrealistically stretched regions.}
\label{fig:qualitative-complete-stretching}
\end{figure}

Quantitative results in Tab.~\ref{tab:resultats} show that our approaches stand out in metrics related to the field smoothness, emphasizing improved regularity in the generated transformations. 
Fig.~\ref{fig:qualitative-complete-stretching} offers a qualitative analysis of the impact of displacement field regularity.
Our method provides smooth displacement fields even for large deformations. 
In contrast, the transformation induced by VoxelMorph, applied directly to the segmentation masks, stretches the liver contours (red arrows). 
Although our approaches operate at the segmentation level rather than the image level, the content-based similarity metrics (NCC and MI) do not show significant differences between all the methods.

Results in Tab.~\ref{tab:results-eval} show that content-based registration methods slightly outperform other methods in matching tumors and achieving higher inclusion ratios for registered tumors.
However, these methods exhibit significantly lower regularity in displacement fields and risk warping lesions to better fit tumors at time t + 1. 
This increases inclusion ratios, but preserving tumor burden alone is insufficient; maintaining tumor shape and appearance is also important for tumor progression evaluation when using both warped and reference images.

\begin{table}[ht]
 \begin{center}
   \tabcolsep = 1\tabcolsep
     \resizebox{0.98\textwidth}{!}{
   \begin{tabular}{l|cccccc}
   \hline\hline
    Method & DSC $\uparrow$ & MI $\uparrow$ & NCC $\uparrow$ & $||\nabla\phi||_2^2$ $\downarrow$ & $||A - A'||_1$ $\downarrow$ & $|J \leq 0|$ $\downarrow$\\
   \hline
    Nifty Reg & 0.96 (0.002) & 0.43 (0.003) & 0.85 (0.002) & 0.09 (0.003) & 0.01 (0.001) & 45145 (4674) \\
    VoxelMorph (NCC + DSC) & 0.98 (0.004) & 0.37 (0.005) & 0.85 (0.002) & 0.04 (0.004) & 0.027 (0.003) & 12845 (1240) 
    \\
    VoxelMorph (DSC) & 0.99 (1e-4) & \textbf{0.457} (0.003) & \textbf{0.89} (0.001) & 0.024 (0.001) & 0.017 (0.002) & 1462 (770) 
    \\
    Diffeomorphic & 0.99 (1e-3) & 0.453 (0.004) & 0.88 (0.001) & 0.006 (0.001) & 0.008 (0.001) & 10 (7) \\
    \hline
    Diffeo\_inc-2& 0.99 (0.001) & 0.454 (0.004) & \textbf{0.89} (0.002) & 0.003 (0.001) & 0.009 (0.001) & 4 (3) 
    \\
    DiffeoCyc\_inc-1 & 0.989 (0.001) & 0.451 (0.003) & 0.884 (0.002) & 0.004 (0.001) & \textbf{0.007} (0.001) & \textbf{0} (0) 
    \\
    DiffeoCyc\_inc-2 & 0.99 (0.001) & 0.453 (0.002) & 0.885 (0.001) & \textbf{0.002} (0.001) & 0.008 (0.001) & 1 (1) \\
   \hline
   \end{tabular}
}
\caption{\textbf{Quantitative results} from longitudinal liver examinations of 33 patients. \textbf{inc-i} represents the number of displacement fields in the forward/backward path.} \label{tab:resultats}
 \end{center}
\end{table}

\begin{table}[ht]
	\begin{center}
		\tabcolsep = 1\tabcolsep
		\resizebox{0.98\textwidth}{!}{
			\begin{tabular}{|l|c|ccc|cc|cc|}
				\hline
				\multirow{2}{2.6cm}& \multirow{2}{2.6cm}{Method Name} & \multirow{2}{2.8cm}{Matched tumors} & \multirow{2}{2.8cm}{Tumor inclusion \\ \hspace{26pt} ratio $\uparrow$} & \multirow{2}{2.8cm}{Tumor burden relative error $\downarrow$} & \multirow{2}{2cm}{$||\nabla\phi||_2^2$ $\downarrow$} & \multirow{2}{2cm}{$|J \leq 0|$ $\downarrow$} & \multirow{2}{1cm}{MI $\uparrow$} & \multirow{2}{1.2cm}{NCC $\uparrow$}\\
				& & & & & & & & \\
				\hline
				Bbox & - & 18/30 & 0.3 & - & - & - & - & - \\ \hline
				Bbox + Affine & - & 22/30 & 0.4 & - & - & - & - & - \\ \hline
				\multirow{7}{2.6cm}{BBox + Affine + Non-Affine} & NiftyReg & \textbf{26}/30  & \textbf{0.58} & 0.162 & 0.08 (0.04) & 34553 (22345) & 0.48 & 0.828 \\ 
				& Vxm (NCC+DSC) & 25/30 & 0.47 & 0.145 & 0.012 (0.002) & 361 (706) & 0.5 & 0.865 \\ 
				& Vxm (DSC) & 25/30 & 0.46 & 0.268 & 0.023 (0.003) & 1000 (1310) & 0.5 & 0.864 \\ 
				& Difféomorphe & 25/30 & 0.45 & 0.137 & 0.005 (0.002) & 12 (46) & 0.5 & 0.86 \\ \cline{2-9}
				& Difféo\_inc-2 & 25/30 & 0.45 & 0.138 & 0.003 (0.002) & 10 (86) & 0.5 & 0.861 \\
				& DifféoCyc\_inc-1 & 25/30 & 0.45 & 0.115 & 0.004 (0.002) & 3 (8) & 0.5 & 0.858\\ 
				& DifféoCyc\_inc-2 & 25/30 & 0.46 & 0.118 & \textbf{0.002} (0.001) & \textbf{0} (0) & 0.5 & 0.86 \\ 
				\bottomrule
			\end{tabular}
		}
		\caption{\textbf{Tumor related quantitative results} for 20 liver cancer patients.} 
		\label{tab:results-eval}
	\end{center}
\end{table}

\paragraph{\textbf{Impact on tumors}}
We consider three cases highlighting the importance of field regularity. Fig.~\ref{fig:qualitative-tumor-single-case} illustrates the case (a) (a detailed illustration of all the cases is in Appendix.~\ref{appendix:qualitative}).

\begin{figure}[h!]
\includegraphics[width=1.0\textwidth]
{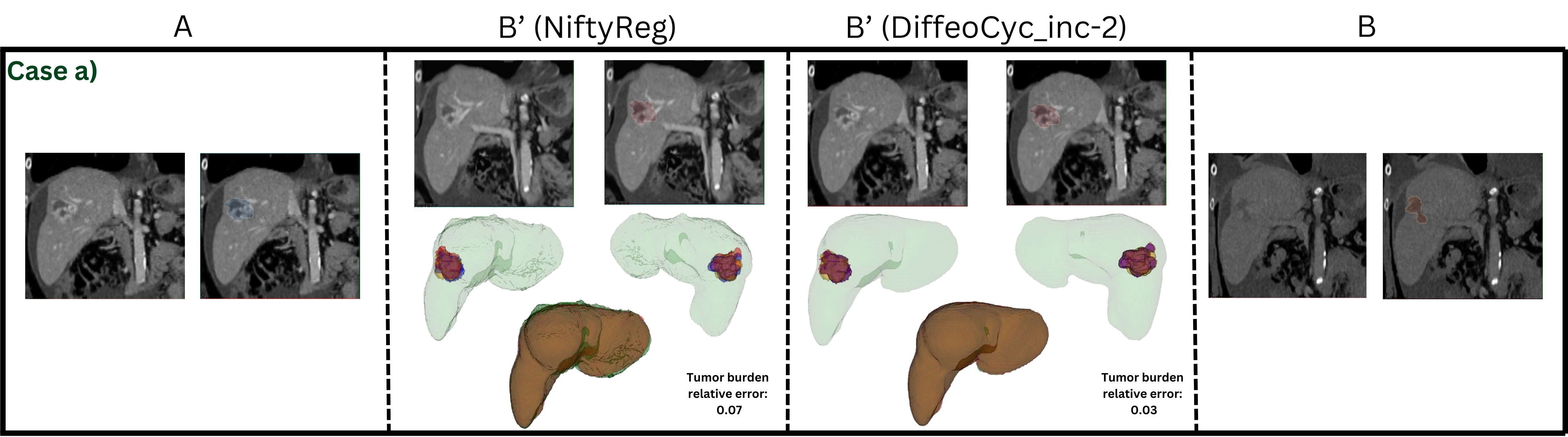}
\centering
\caption{Left to right: Moving image A (tumor in blue), transformed image B' for \textit{NiftyReg} and our proposed framework \textit{DiffeoCyc\_inc-2} (tumor in red), and fixed image B (tumor in orange). The transformed liver masks are represented in green, and the fixed image mask B is represented in red.}
\label{fig:qualitative-tumor-single-case}
\end{figure}

\noindent \textbf{\(\star \) Case (a):} We examine a benign tumor where both NiftyReg and DiffeoCyc\_inc-2  yield comparable relative errors. However, how alterations are applied differs significantly. NiftyReg introduces “stretching” in some tumor areas, whereas DiffeoCyc\_inc-2 ensures a more uniform deformation. It is essential to recognize that even when relative errors are similar, the specific deformation strategy can impact diameter measurements, which is the criteria used for tumor progression assessment according to RECIST \cite{RECIST}. For instance, in this case, a relative error of 0.06 corresponds to a volumetric discrepancy of 2.268 mL (2268 mm³). Given that clinical workflows demand liver tumor diameter measurements accurate up to 5 mm, such nuances become crucial.

\noindent \textbf{\(\star \) Cases (b) and c):} In these scenarios, the tumor in the moving image (blue) grows with time in the fixed image (orange). The transformation applied to the tumor (red) using the NiftyReg method stretches the tumor to match the larger tumor in the fixed image, which is larger. Achieving this alignment comes at the cost of a slightly elevated relative error due to warping during registration, highlighting the trade-off between alignment accuracy and error.  This also explains the high tumor inclusion ratio in content-based approaches.

The observed results emphasize the role of smooth deformations within the context of registration when the model treats both warped and fixed images. When assessing tumor progression after registration, ensuring that the tumor burden remains relatively stable is not the sole consideration, as tumors can undergo localized deformations while maintaining overall volume. However, such localized deformations may introduce measurement bias, particularly when assessing diameter or other quantitative metrics. Therefore, evaluating registration methods based on a combination of tumor burden conservation and displacement field smoothness provides a more robust metric in this context.

\section{Conclusion}
In this work, we introduced a registration framework designed to assist tumor progression assessment in longitudinal liver CT scans. 
Leveraging an automatic segmentation model, our framework aligns liver segmentation masks through smooth displacement fields. This alignment targets tumor alignment with minimal tumor deformation. Even though the internal structures of the liver may not be perfectly aligned after registration, their shape will be nearly preserved while being brought much closer to their counterpart in the reference image. The impact of this registration on a subsequent tumor progress/change detection module is to be addressed in future works.

\begin{credits}
\subsubsection{\ackname} This preprint has not undergone peer review or any post-submission improvements or corrections. The Version of Record of this contribution is published in Medical Image Computing and Computer Assisted Intervention
– MICCAI 2024 Workshops ( Springer Nature Switzerland, 2025), to be published in March 2025.

\end{credits}

\newpage

\bibliographystyle{splncs04}
\bibliography{biblio}

\appendix

\section{Bidirectional vs Cyclic Approach}\label{appendix:cyclic}

\textbf{Why is the inverse consistency loss not well defined for a bidirectional approach?} Let $b'$ and $a'$ be the new positions of $a$ and $b$ after registration by $\phi_{AB}$ and $\phi_{BA}$, respectively. 
As long as $a' \neq a$ and $b' \neq b$, the inverse consistency loss can not be zero, $\forall a \in S_A$ and $\forall b \in S_B$. While we impose the fields $\phi_{a->b'}$ and $\phi_{b->a'}$ to be inversely consistent, the starting point ($b$) of the second field does not necessarily coincide with the ending point ($b'$) of the first. 
The following diagram illustrates the motivation behind employing a cyclic approach rather than a bidirectional one, as presented in section \ref{Methodology}, paragraph \ref{paragraph:inverse-coherence}.

\begin{figure}[h!]
\includegraphics[width=0.7\textwidth]{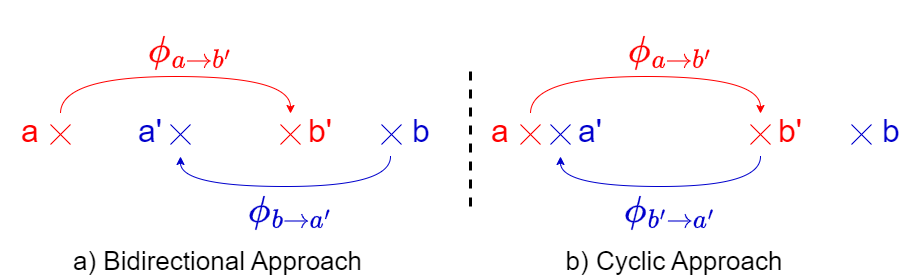}
\centering
\caption{Bidirectional (a) vs. cyclic (b) approach: Diagram illustrating how the displacement fields are defined in each approach.}
\label{fig:methodology-diagram}
\end{figure}

\section{Public Datasets}\label{appendix:literature}

In the healthcare domain, releasing public datasets involves a rigorous process due to the sensitive nature of the information they contain. 
Longitudinal studies further complicate this process, requiring patients to undergo at least two examinations at specific time intervals. This scarcity of public longitudinal datasets is particularly evident in our case for abdominal CT scans of the liver. 
Existing public datasets, such as ADNI-2 \cite{ADNI-2}, ISBI-2015 (MS lesion segmentation challenge by  \cite{Longitudinal_MS}), OASIS-2 \cite{OASIS-2} and OASIS-3 \cite{OASIS-3}, primarily focus on brain studies, making them unsuitable for our specific use case. 
This is attributed to the inherent disparity between the longitudinal brain and abdominal images, as discussed in Section \ref{Introduction}, especially since the impact of the temporal dimension is more pronounced in abdominal images than in brain images.
Learn2Reg \cite{Learn2Reg} offers data for CT-MRI modalities in inter-patient or intra-patient abdominal examinations. However, it lacks the essential longitudinal aspect needed to address the temporal nature of our specific use case.

\newpage

\section{Qualitative Results: Impact on tumors}\label{appendix:qualitative}

These masks represent three-dimensional (3D) delineations of tumors, color-coded as follows: blue for the tumors in the moving image, red for tumors in the transformed image, and orange for tumors in the fixed image. Multiple 3D views of tumors with the transformed liver segmentation mask are provided to facilitate visualizing the 3D structures. An additional 3D overlay is presented for the transformed liver mask (in green) and the fixed image liver mask (in red).

\begin{figure}[h!]
\includegraphics[width=0.99\textwidth]
{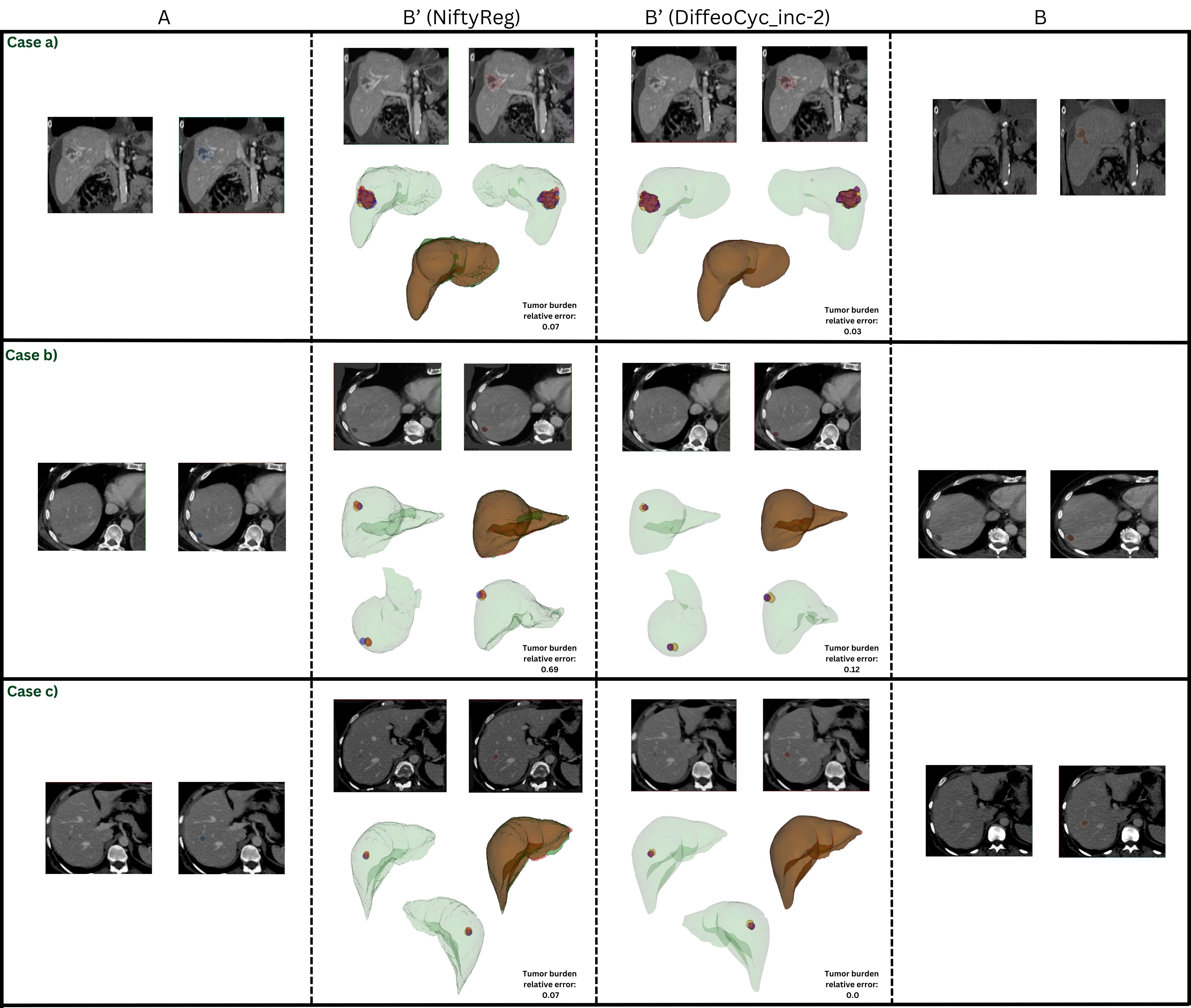}
\centering
\caption{Left to right: Moving image A (tumor in blue), transformed image B' for \textit{NiftyReg} and our proposed framework \textit{DiffeoCyc\_inc-2} (tumor in red), and fixed image B (tumor in orange). The transformed liver masks are represented in green, and the fixed image mask B is represented in red.}
\label{fig:qualitative-tumor}
\end{figure}

\end{document}